# Chiral damping of magnetic domain walls

Emilie Jué[1,2,3]†, C.K. Safeer[1,2,3], Marc Drouard[1,2,3], Alexandre Lopez[1,2,3], Paul Balint[1,2,3],

Liliana Buda-Prejbeanu[1,2,3], Olivier Boulle[1,2,3], Stephane Auffret[1,2,3], Alain Schuhl[4],

Aurelien Manchon[5], Ioan Mihai Miron[1,2,3]∗, & Gilles Gaudin[1,2,3]

[1]*Univ. Grenoble Alpes INAC-SPINTEC F-38000 Grenoble France*

[2]*CNRS INAC-SPINTEC F-38000 Grenoble France*

[3]*CEA INAC-SPINTEC F-38000 Grenoble France*

[4]*CNRS Institut Néel Grenoble France*

[5]*King Abdullah University of Science and Technology (KAUST), Physical Science and Engineering Division, Thuwal 23955-6900, Saudi Arabia*

\* To whom correspondence should be addressed.
E-mail: mihai.miron@cea.fr

† Currently at NIST Boulder, Colorado, US

**Structural symmetry breaking in magnetic materials is responsible for a variety of outstanding physical phenomena. Examples range from the existence of multiferroics[1], to current induced spin orbit torques (SOT)[2-7] and the formation of topological magnetic structures[8-11]. In this letter we bring into light a novel effect of the structural inversion asymmetry (SIA): a chiral damping mechanism. This phenomenon is evidenced by measuring the field driven domain wall (DW) motion in perpendicularly magnetized asymmetric Pt/Co/Pt trilayers. The difficulty in evidencing the chiral damping is that the ensuing DW dynamics exhibit identical spatial symmetry to those expected from the Dzyaloshinskii-Moriya interaction (DMI)[12-18]. Despite this fundamental resemblance, the two scenarios are differentiated by their time reversal properties: while DMI is a conservative effect that can be modeled by an effective field, the chiral damping is purely dissipative and has no influence on the equilibrium magnetic texture. When the DW motion is modulated by an in-plane magnetic field, it reveals the structure of the internal fields experienced by the DWs, allowing to distinguish the physical mechanism. The observation of the chiral damping, not only enriches the spectrum of physical phenomena engendered by the SIA, but since it can coexists with DMI it is essential for conceiving DW and skyrmion devices[19].**

The role of the SIA on the magnetic properties of the metallic ferromagnets has been the focus of extensive research. Whether the SIA occurs at an interface or in the bulk of a material, it may induce non-collinear exchange interactions[8] (DMI). This gives rise to a rich variety of magnetic phases, such as spin helices[9,10] and skyrmions[11] or to magnetic frustration[23]. Recently, it was proposed that the DMI promotes the existence of Néel DWs in materials where the configuration otherwise favored would be Bloch[12]. In the context of current induced DW motion[20-22] this is an important idea, as it may solve a long-standing issue of DWs moving in the direction of the electric current and not into that of the electron flow, as one would expect from spin transfer torque[12-14]. Moreover, the observation that the field-driven DW motion in Pt/Co based multilayers is made asymmetric by in-plane magnetic fields was also linked to the possible occurrence of DMI[15,16].

Until now, DMI was the only well established interaction known to link the DW magnetic texture to the SIA of the layer structure. In this letter we propose a new mechanism, where the magnetic texture can sense the SIA as a damping whose intensity depends on the DW chirality.

DWs in materials with perpendicular magnetic anisotropy (PMA) are the perfect test-ground for such phenomena: they are chiral objects and their motion is relatively easy to detect. From the perspective of the magnetization dynamics, all relevant physical phenomena translate into torques that control the magnetization according to the phenomenological Landau–Lifshitz–Gilbert (LLG) equation[24]. Their time reversal properties separate them in two categories. They can, either derive from a form of energy and be invariant to time reversal (anisotropy, exchange), or they may be dissipative and break time reversal symmetry (damping). These two scenarios are exclusive, in the sense that there is no other possibility. Similarly, the physical phenomenon linking the DW motion asymmetry to the material's SIA must fall into one of these categories. It can only derive from chiral energy (DMI) or be a form of chiral dissipation.

The effect of chiral energy has already been evidenced experimentally[15]. The DMI implies the existence of an internal field $H_{DMI}$ setting the DW chirality (Figure 1b). As a consequence, the core magnetization of up/down (↑↓) and down/up (↓↑) DWs will point in opposite directions. The field driven DW velocity can sense this interaction if an external in-plane field ($H_{ip}$) is applied simultaneously with the perpendicular field ($H_z$). ↑↓ and ↓↑ DWs moving along $H_{ip}$ will experience different effective fields ($H_{ip}$-$H_{DMI}$ and $H_{ip}$+$H_{DMI}$) and therefore will have different velocity. As long as the applied field does not alter the equilibrium

DW profile (by significantly tilting the domains) the DW dynamics are correctly described by a collective coordinate model. Since in this case $H_{ip}$ and $H_{DMI}$ enter the DW equation of motion on equal footing[15], DMI will bias the velocity dependence on $H_{ip}$ by $+H_{DMI}$ and $-H_{DMI}$ respectively (Figure 1b), making the motion asymmetric.

The chiral damping mechanism, although allowed by symmetry has never been observed experimentally. Its chief attribute is that, unlike DMI, it causes asymmetric displacements without biasing the magnetic configuration (Figure 1d).

We use wide field Kerr microscopy to probe the DW motion in asymmetric $Pt_{30Å}/Co_{6Å}/Pt_{xÅ}$ trilayers subjected simultaneously to $H_z$ and $H_{ip}$. A pulse of $H_z$ nucleates a reversed domain. A second pulse of $H_z$ accompanied by $H_{ip}$ further propagates the circular DW. Figure 2a shows typical magnetic differential images corresponding to the DW displacement produced by $H_z$ and $H_{ip}$. The reversed magnetization produces bright (or dark) contrast. Without $H_{ip}$ the DWs move over equal distances in all directions. The direction and the magnitude of the asymmetry are controlled by $H_{ip}$. This is qualitatively consistent with the existence of the $H_{DMI}$: depending on the type of DW, (↑↓) or (↓↑), $H_{ip}$ either adds-up or subtracts to $H_{DMI}$, leading to different effective field for the two DWs. To verify this possibility, we measured the DW velocity ($v_{DW}$) vs. $H_{ip}$ for a fixed $H_z$. While $H_{ip}$ can be applied continuously, as it provokes no DW motion, $H_z$ is applied in pulses of 400 ms. $v_{DW}$ is extracted by dividing the displacement observed on the Kerr image by the pulse length.

Figure 2 shows the effect of $H_{ip}$ on $v_{DW}$. Besides the chiral effects emerging from SIA, $H_{ip}$ may also affect $v_{DW}$ through phenomena unrelated to SIA (such as the variations of the effective anisotropy). To separate the chiral effects we decompose $v_{DW}$ into a symmetric (*S*) and anti-symmetric (*A*) component with respect to $H_{ip}$: $v_{\uparrow\downarrow} = S(1+A/2)$ and $v_{\downarrow\uparrow} = S(1-A/2)$, where $S = (v_{\uparrow\downarrow} - v_{\downarrow\uparrow})/2$ and $A = 2 \cdot (v_{\uparrow\downarrow} - v_{\downarrow\uparrow})/(v_{\uparrow\downarrow} + v_{\downarrow\uparrow})$. While *S* can be affected by non-SIA phenomena, *A* is a robust indicator of chiral effects, as by symmetry, only SIA related phenomena can create it. Since *A* dictates the ratio $v_{\uparrow\downarrow}/v_{\downarrow\uparrow} = \frac{A+2}{2-A}$, it is also an indicator of the asymmetry of DW motion the along the direction of $H_{ip}$ (Figure 2a).

The variation of *A* vs. $H_{ip}$ has two distinctive features: it varies continuously up to a saturation field ($H_{sat}$=40mT), and then saturates (Figure 2a,2b), meaning that below $H_{sat}$ the velocities of the two DWs vary differently, while above $H_{sat}$ they vary identically to each other. This indicates that the parameter responsible for $v_{\uparrow\downarrow} \neq v_{\downarrow\uparrow}$ must also saturate at $H_{sat}$. In our samples, the only micromagnetic parameter that may vary monotonically up to 40 mT and then saturate, is the DW core magnetization. Due to the large uniaxial anisotropy ($H_k$=500 mT)[25], the domain tilting affecting the DW internal structure and the effective DW width, becomes significant at much higher fields.

$H_{sat}$ must then designate the field value for which both DWs become magnetized along $H_{ip}$, implying that $H_{DMI}$ is smaller than $H_{sat}$. If DMI imposes homochiral Néel DWs, $H_{ip}$ acts only on the magnetization of one of the two DWs, as the other one is already aligned (Figure 1a). Therefore, $v_{DW}$ should display a feature at $H_{sat}$ for only one of the DWs (Figure 1c). On the contrary, we observe (Figure 2c,S5) that $v_{DW}$ vs. $H_{ip}$ changes slope close to $H_{sat}$ for both DWs. The fact that $H_{sat}$ is physically relevant for both DWs excludes the existence of homochiral Néel DWs in our samples. It also sets a ceiling value for $H_{DMI}$ within the precision of $H_{sat}$ determination (±10mT).

As explained earlier, if *A* is produced by a small DMI (that we cannot rule out completely), the velocity curves for the two DWs should have the same shape and only differ through a horizontal offset given by the value of the $H_{DMI}$ bias field (Figure 1b). By shifting the two curves in opposite directions by $H_{DMI}$, they should coincide. Clearly, the measured $v_{DW}$ do not follow this scenario: it is impossible to overlap the curves by any amount of lateral shifting (Figure 2c).

Based on the variations of $v_{DW}$ induced by $H_{ip}$ at constant $H_z$, we established that our samples support Bloch DWs and that the asymmetry mechanism evidenced in samples with Néel DWs[15,16] does not apply here. Now we switch perspectives and study the DW motion as a function of $H_z$ at fixed $H_{ip}$. To separate the effect of $H_{ip}$ on the chiral energy and chiral damping, we rely on the well-known creep scaling law, describing the DW velocity at very small driving fields:

$$v_{creep} = v_0 \cdot \exp\left[-\frac{U_c H_p^{1/4}}{k_B T} \cdot H_z^{-1/4}\right] \quad (1)$$

$U_c$ - height of the pinning barrier; $H_p$ - pinning field;

Since the prefactor[31] $v_0 = d_0 f_0 \cdot \exp\left[C \cdot \frac{U_c}{k_B T}\right]$ ($C \approx 1$ is an empirical constant) also depends exponentially on the pinning barrier, it is more convenient to write the creep scaling law[15,26] as

$$v_{DW} = d_0 f_0 \cdot \exp\left[-\frac{E}{k_B T} \cdot \left(H_z^{-1/4} - H_c^{-1/4}\right)\right] \quad (2)$$

and group all the terms related to energy into a single exponential ($E = U_c H_p^{1/4}$). $H_c = \frac{H_p}{C^4}$ is the critical field that determines the limit of the creep regime[31]. Since the dissipation does not contribute to the exponent, the damping can only affect the prefactor[25-26] $d_0 f_0$ [31] ($d_0$ – disorder correlation length; $f_0$ – the attempt frequency).

As long as the creep scaling law applies, the plot of $\ln(v_{DW}) vs. H_z^{-1/4}$ allows to analyze separately the effect of $H_{ip}$ on the prefactor and the exponent, as they appear as the intercept and the slope of the linear dependence (Figure S2). Figure 3a,b shows typical results for two samples with different capping thickness ($Pt_{15.6Å}$ and $Pt_{30Å}$). The linear dependence specific for the creep regime is observed for the two samples and both DW polarities. To evidence the symmetric and anti-symmetric effects of $H_{ip}$ we consider

$$A_{creep} = \ln(v^{80mT \uparrow\downarrow}) - \ln(v^{80mT \downarrow\uparrow}) \qquad (3)$$

$$S_{creep} = \frac{1}{2}\left[\ln(v^{80mT \uparrow\downarrow}) + \ln(v^{80mT \downarrow\uparrow})\right] - \ln(v^{0mT}) \qquad (4)$$

According to (2) they can be written

$$A_{creep} = \ln\left(\frac{d_0 f_0^{80mT \uparrow\downarrow}}{d_0 f_0^{80mT \downarrow\uparrow}}\right) - \left(H_z^{-1/4} - H_c^{-1/4}\right)\left(\frac{E^{80mT \uparrow\downarrow} - E^{80mT \downarrow\uparrow}}{k_B T}\right) \qquad (5)$$

$$S_{creep} = \ln\left(\frac{\sqrt{d_0 f_0^{80mT \uparrow\downarrow} \cdot d_0 f_0^{80mT \downarrow\uparrow}}}{d_0 f_0^{0mT}}\right) - \left(H_z^{-1/4} - H_c^{-1/4}\right)\left(\frac{\frac{1}{2}\left(E^{80mT \uparrow\downarrow} + E^{80mT \downarrow\uparrow}\right) - E^{0mT}}{k_B T}\right) \qquad (6)$$

Remarkably, $A_{creep}$ shows negligible dependence on $H_z^{-1/4}$ (inset of Figure 3a,b), implying that the bubble asymmetry is not due to chiral energy ($E^{80mT \uparrow\downarrow} = E^{80mT \downarrow\uparrow}$) but to chiral dissipation ($d_0 f_0^{80mT \uparrow\downarrow} \neq d_0 f_0^{80mT \downarrow\uparrow}$). At the same time, $S_{creep}$ decreases linearly with $H_z^{-1/4}$, signaling the existence of non-chiral energy contributions related to $H_{ip}$ ($E^{80mT \uparrow\downarrow} = E^{80mT \downarrow\uparrow} \neq E^{0mT}$).

Though it has been predicted[27] that friction affects the creep of an elastic membrane via the velocity prefactor, to our knowledge, in the case of DWs no microscopic mechanism has yet been proposed. For this particular case, the physical parameter most likely to depend on damping is the attempt frequency $f_0$. We verify this possibility using micromagnetic simulations by calculating the DW depinning time from a single defect for different damping values (Figure 3c,d). Indeed, we observe a strong dependence of the release time on damping. Since the barrier height is kept constant in the simulations, it is only $f_0$ that can be responsible for this variation. We find that $f_0$ is proportional to the inverse of the damping (DW mobility). This trend ($v_{DW} = 1/\alpha$) is further confirmed for a wider range of applied fields and damping values using a 1D model[25].

The above mechanism can account for the chiral DW dynamics if the damping coefficient depends on the orientation of the DW's core with respect to the magnetization gradient: $\alpha \propto \alpha_C (\vec{m}_{ip} \cdot \nabla m_Z)$. However, this kind of damping cannot exist by itself in a real physical system since the dissipation may become negative if the chiral contribution exceeds the intrinsic damping. To prevent this unphysical behavior, the chiral damping must include at least a second component, which always offsets its value in the positive range. For example, its mathematical expression could be $\alpha \propto \alpha_0 + \alpha_C (\vec{m}_{ip} \cdot \nabla m_Z)$, where $\alpha_0 > \alpha_C$. Based on this expression, the amplitude of the chiral damping contribution can be extracted from the value of $A$. Above $H_{sat}$, where $m_{ip}$ saturates, $\frac{v_{\uparrow\downarrow}}{v_{\downarrow\uparrow}} = \frac{\alpha_0 + \alpha_c}{\alpha_0 - \alpha_c}$ and $A/2 = \frac{v_{\uparrow\downarrow}/v_{\downarrow\uparrow} - 1}{v_{\uparrow\downarrow}/v_{\downarrow\uparrow} + 1}$ we obtain $A/2 = \frac{\alpha_c}{\alpha_0}$

For the sample exhibiting the largest asymmetry (Figure S1), the amplitude of the chiral damping reaches approximately 50% of the constant part.

To illustrate the effect of the chiral damping on the DW dynamics we use a numerical collective coordinate model. Since such models do not include explicitly the magnetization gradients, we write the chiral damping using a simplified form: $\alpha = \alpha_0 + \alpha_C \cos\phi$. Here the value of the magnetization gradient is considered constant and is implicitly included in the values of $\alpha_0$ and $\alpha_C$. The orientation of the DW magnetization is described by the azimuthal angle $\phi$. Figure 2e shows the computed $v_{DW}$ as a function of $H_{ip}$ at 0K and without pinning. The salient feature is that the velocities of both DWs saturate simultaneously, at the field value where their core magnetization saturates. In order to compare the experimental results with the simulations, the symmetric component $S$ of the measured velocity must be removed, as it is not produced by the chiral damping. For this purpose we normalize $v_{\downarrow\uparrow}$ and $v_{\uparrow\downarrow}$ to the average velocity along $x$ and $y$, the in-plane directions. The normalized DW velocity (Figure 2d, Figure S5) saturates at $H_{sat}$, in perfect agreement with the simulations.

At this stage, we do not understand why, contrary to previous studies, the effect of the chiral damping in our samples is larger than the effect of DMI. We infer that these two phenomena should co-exist, but their relative strength may be strongly dependent on the precise crystallographic structure. An independent hint of the co-existence of chiral damping and DMI was provided by the magnon dynamics in Fe/W bi-layers, where it was observed that the not only the dispersion relation was shifted due to DMI, but also that magnon lifetime depends on the chirality[28].

While theoretical work is definitely required to understand the origin of the chiral damping, we would like to speculate and point towards a plausible explanation: both theoretically and experimentally it was shown that the spin orbit torques have two components: a field like[2,3] (conservative) part and a damping like (dissipative) part[4,5,6]. Theoretically the DMI was linked to the conservative (field like) part[29,30], as they share a common microscopic origin. In a similar way, the chiral damping reported here, might be reminiscent of its dissipative component.

From a practical perspective, the DW motion asymmetry can be useful for memory[22] or logic devices[32] requiring unidirectional DW movement. The application of an alternating $H_z$ causes the domain to enlarge and then shrink to its initial position. When $H_{ip}$ is applied along with $H_z$, the DW motion will become asymmetric. If $H_{ip}$ changes sign together with $H_z$, the asymmetry of the domain growth will be opposite to that of its contraction. Consequently, the domain will not come back to its initial position but will be shifted along the axis of the in-plane field[15]. The relative sign of the two field components controls the direction. Compared to current induced DW motion where trains of DWs are pushed along 1D wires, the phenomenon that we describe affords a distinctive ability of shifting 2D magnetic domains in any directions of a full sheet magnetic layer (Figure 4c; supplementary movie).

In conclusion, the quantitative analysis of field induced DW motion has revealed the existence of a chiral damping term whose effects on the DW dynamics have the same spatial symmetry as those expected from the DMI. Understanding and controlling the chiral damping is essential for the designing spintronics devices that rely on chiral magnetic textures, such as DW or skyrmion racetracks.

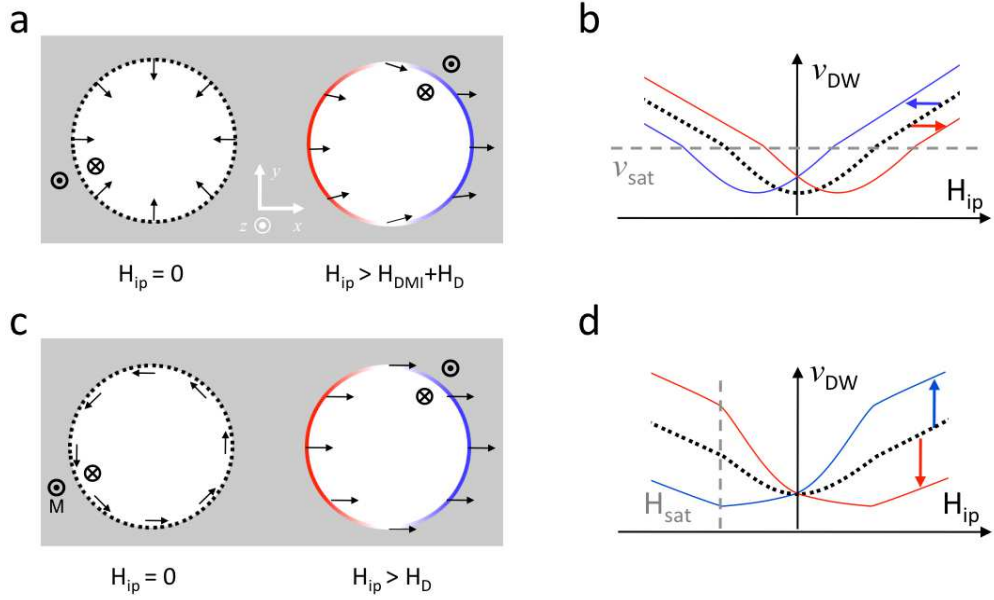

**Figure 1.** Graphic illustration of the asymmetric DW dynamics in PMA materials. **a.** A DW (dotted line) separates two oppositely magnetized domains (grey and white). The DMI can be modeled by an effective field that only exists inside the DW and that always rotates the DW magnetization to point from the domain pointing up to the one pointing down. If $H_{ip}$ is applied along *x*, the two DWs standing orthogonal to its direction will react differently: while one remains magnetized in the same direction (red), the other will reverse its chirality (blue). **b** Without SIA and hence with no DMI to differentiate the two DWs, $H_{ip}$ will have the same effect on the field driven velocity of the two DWs that move along *x* (black dashed lines). If present, the $H_{DMI}$ bias field will shift laterally the velocity curves for the two DWs (red and blue lines). An important consequence is that any feature of the initial velocity curve allowing to identify the DW transformation, such as a change of slope when its magnetization saturates, will translate laterally, and always manifest at the same velocity ($v_{sat}$). **c.** If no DMI is present, the DWs have Bloch structure; the structural changes produced by $H_{ip}$ will be identical for the two DWs. **d.** The chiral damping does not shift laterally the initial "non-SIA" velocity curve (black dotted line), but creates an asymmetric vertical correction. In this case, the DW transformation from Bloch to Néel, will not occur at the same velocity, but at the same field value ($H_{sat}$).

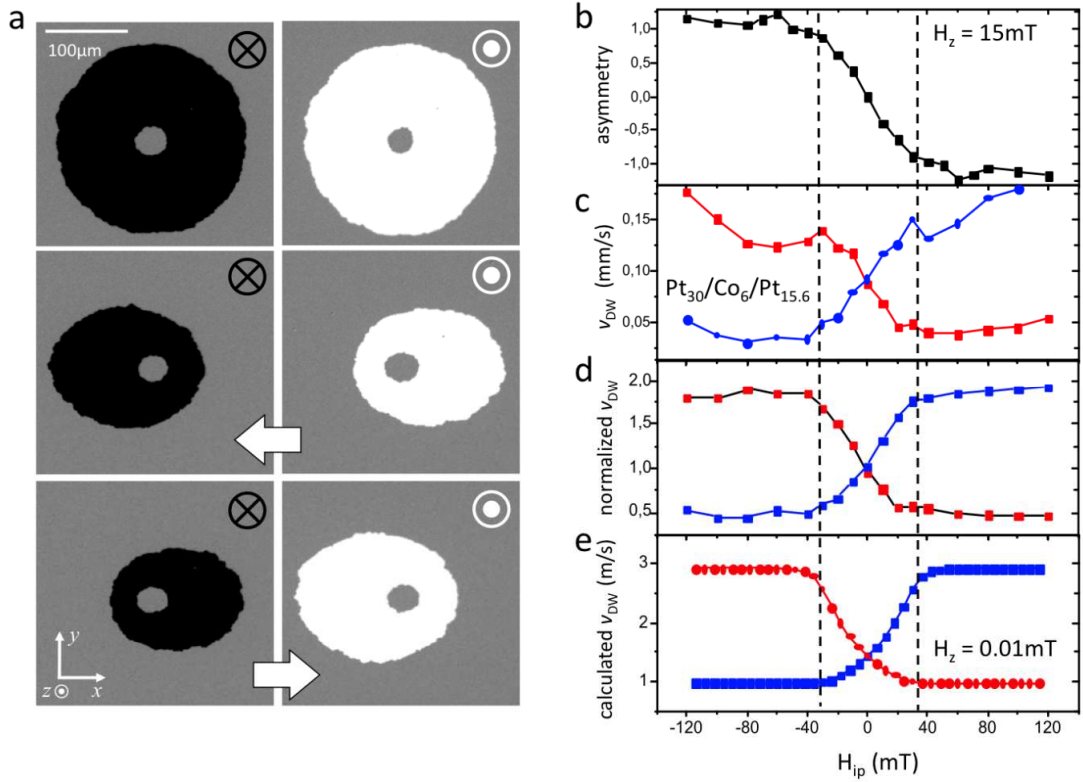

**Figure 2**. DW velocity modulation by $H_{ip}$ in Pt/Co/Pt layers. **a.** Kerr microscopy images of field induced DW displacements. Dark contrast corresponds to the growth of a "down" domain; Bright contrast corresponds to an "up" domain. Without in-plane field the DW motion is isotropic. The in-plane field (white arrows) makes the bubble growth asymmetric: in the field direction for an "up" domain; opposite to the field direction for a "down" domain. It appears that the asymmetry depends on the direction of the in-plane component of the DW's magnetization ($\vec{m}_{ip}$) with respect to the gradient of its vertical component ($\nabla m_z$), defining a chirality. **b.** The anti-symmetric component of the DW velocity ($Pt_{30}/Co_6/Pt_{15.6}$ sample) varies monotonically and then saturates at a characteristic field value ($H_{sat}$). Since any deviations from the condition: $v\downarrow\uparrow(H_{ip}) = v\uparrow\downarrow(-H_{ip})$, are caused by noise or measurement artifacts, to improve the signal/noise ratio we plot $A = \frac{1}{2}(A_{raw}(H_{ip}) - A_{raw}(-H_{ip}))$, instead of $A_{raw} = 2(v\downarrow\uparrow - v\uparrow\downarrow)/(v\downarrow\uparrow + v\uparrow\downarrow)$ (Figure S5) **c.** The measured DW velocity exhibits changes in the slope at $H_{sat}$ for both DWs (red and blue) and both directions of $H_{ip}$. Note that despite the different overall shape of the DW velocity dependence for the two samples, both measurements exhibit a change of slope

at H$_{sat}$ **d.** the velocity of the left and right moving DWs normalized to the average velocity along the in-plane directions (*x* and *y*) show that while one of the DWs is accelerated the other slows down. Both these relative variations saturate at H$_{sat}$. **e.** DW velocity calculated by a collective coordinate model using a chiral damping defined as $\alpha \propto \alpha_0 + \alpha_C \cdot m_X$. The red and black curves correspond to the ↑↓ and the ↓↑ DWs. The values were used in the calculation are: $\alpha_0$=0.6 $\alpha_c$=0.3, ΔDW=5nm, H$_{dip}$=35mT, H$_z$=0.01mT.

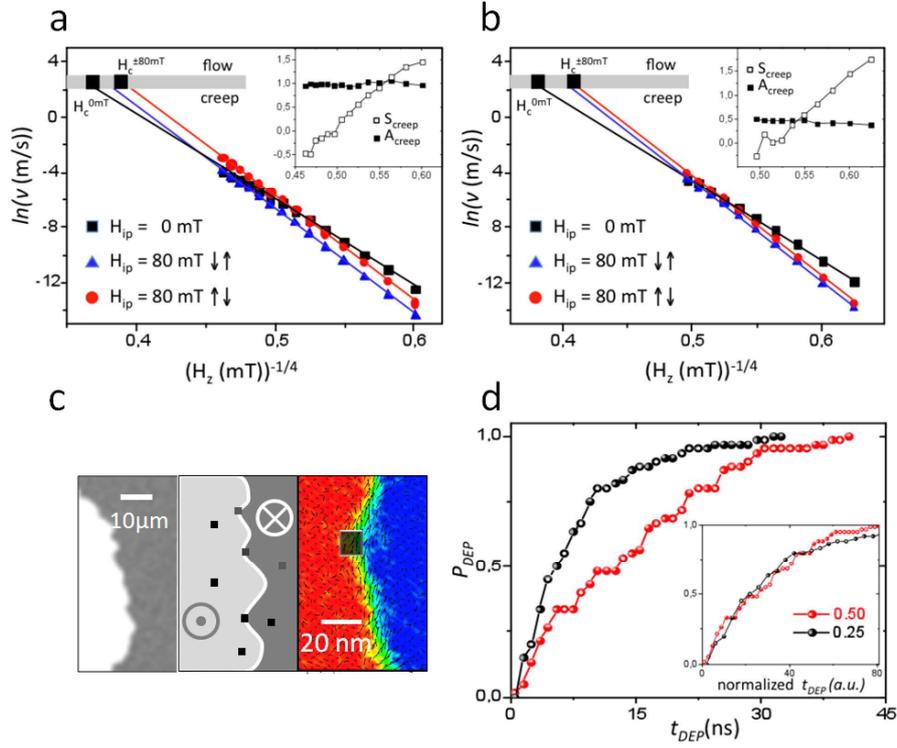

**Figure 3.** Effect of chiral damping on thermally activated DW motion **a,b.** Creep scaling law for $v{\downarrow}{\uparrow}$ and $v{\uparrow}{\downarrow}$ for the $Pt_{30}Co_6Pt_{15.6}$ and $Pt_{30}Co_6Pt_{30}$ samples. The measurements were performed at $H_{ip}=0$ (black) and $H_{ip}=80$ mT (red and blue). The grey horizontal line is the limit of the creep regime, where we expect the transition to the flow (Figure S2). The rectangular black square at the intercept of the linear fit with this limiting region is an approximate indicator of $H_c^{-1/4}$. In the inset we plot $S_{creep}$ (empty symbols) and $A_{creep}$ (full symbols) **c.** Design of the micromagnetic model system. The first picture from left to right is a differential Kerr image of a DW propagated by $H_z$ showing the typical DW shape. The second one is a schematic representation of the localized pinning that creates the rippled DW shape evidenced by Kerr imaging. The third picture is a micromagnetic configuration depicting a pinned DW in a 100 nm wide nanowire pinned to a localized defect; the dark rectangle designates an area with 50% smaller anisotropy that attracts the DW **d.** The simulated depinning probability (60 independent events) for two damping values (0.5 and 0.25) at Hz=18mT. The de-pinning field at 0K is 30mT. The inset shows that the curves overlap when the timescale is normalized by the damping constant. This indicates that $f_0$ is proportional to the inverse of the damping.

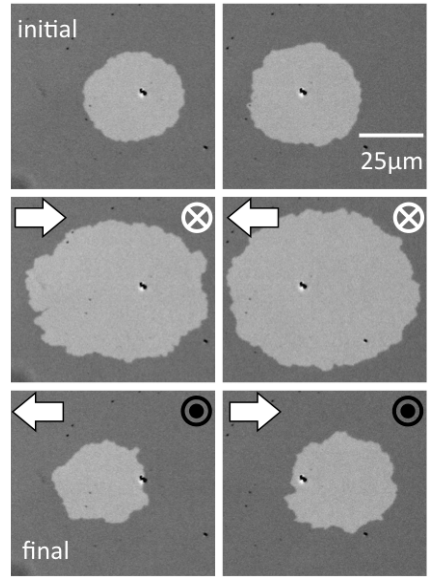

**Figure 4**. Bubble shifting by magnetic fields. Sequence of direct Kerr images showing the bubble shifting in the $Pt_{30}Co_6Pt_{18,7}$ sample by combined in-plane and perpendicular fields ($H_{ip}$ = 40mT and $H_z$ = 11mT) The relative direction of the in-plane and perpendicular magnetic field is indicated on the images. The same experimental parameters are used in the supplementary movie.

References:

1. Khomskii, D. Classifying multiferroics: Mechanisms and effects *Physics* 2, 20 (2009)

2. Chernyshov, A. *et al*. Evidence for reversible control of magnetization in a ferromagnetic material by means of spin–orbit magnetic field. *Nature Phys.* **5**, 656–659 (2009).

3. Miron, I. M. *et al*. Current-driven spin torque induced by the Rashba effect in a ferromagnetic metal layer. *Nature Mater.* **9**, 230–234 (2010).

4. Miron, I. M. *et al*. Perpendicular switching of a single ferromagnetic layer induced by in-plane current injection. *Nature* **476**, 189–193 (2011).

5. Liu, L. *et al*. Spin-torque switching with the giant spin Hall effect of tantalum. *Science* **336**, 555–558 (2012).

6. Fang, D. *et al*. Spin–orbit-driven ferromagnetic resonance. *Nature Nanotech.* **6**, 413–417 (2011).

7. Manchon, A. & Zhang, S. Theory of nonequilibrium intrinsic spin torque in a single nanomagnet. *Phys. Rev. B* **78**, 212405 (2008).

8. Dzyaloshinskii, I. E. Thermodynamic theory of weak ferromagnetism in antiferromagnetic substances. *Sov. Phys. JETP* **5**, 1259–1262 (1957). Moriya, T. Anisotropic superexchange interaction and weak ferromagnetism. *Phys. Rev.* **120**, 91–98 (1960).

9. Bode, M. *et al*. Chiral magnetic order at surfaces driven by inversion asymmetry. *Nature* **447**, 190–193 (2007).

10. Ishikawa, Y., Tajima, K., Bloch, D., & Roth, M. Helical spin structure in manganese silicide MnSi. *Solid State Communications*, *19*(6), 525-528. (1976).

11. Mühlbauer, S. *et al*. Skyrmion lattice in a chiral magnet. *Science* **323**, 915–919 (2009).

12. Thiaville, A., Rohart, S., Jué, É., Cros, V., & Fert, A. Dynamics of Dzyaloshinskii domain walls in ultrathin magnetic films. *EPL (Europhysics Letters)*, *100*(5), 57002. (2012).

13. Ryu, K. S., Thomas, L., Yang, S. H., & Parkin, S. Chiral spin torque at magnetic domain

walls. *Nature Nanotechnology 8(7), 527-533.* (2013).

14. Emori, S., Bauer, U., Ahn, S. M., Martinez, E., & Beach, G. S. Current-driven dynamics of chiral ferromagnetic domain walls. *Nature materials*, *12*(7), 611-616 (2013).

15. Je, S-G et al. Asymmetric Magnetic Domain-Wall Motion by the Dzyaloshinskii-Moriya Interaction Physical Review B 88, 214401 (2013); Moon, Kyoung-Woong, et al. "Magnetic Bubblecade Memory." *arXiv:1409.5610* (2014).

16. Hrabec, A. *et al.*, DMI meter: Measuring the Dzyaloshinskii-Moriya interaction inversion in Pt/Co/Ir/Pt multilayers. *arXiv preprint arXiv:1402.5410.* (2014)

17. Torrejon, J., et al., Interface control of the magnetic chirality in CoFeB| MgO heterosctructures with heavy metal underlayers. *arXiv preprint arXiv:1401.3568* (2014)

18. Chen, G. *et al.*, Tailoring the chirality of magnetic domain walls by interface engineering. *Nature communications*, *4*. 2671 (2013)

19. Sampaio, J., Cros, V., Rohart, S., Thiaville, A., & Fert, A. (2013). Nucleation, stability and current-induced motion of isolated magnetic skyrmions in nanostructures. *Nature nanotechnology*, *8*(11), 839-844 (2013)

20. Berger, L. Exchange interaction between ferromagnetic domain wall and electric current in very thin metallic films. *Journal of Applied Physics*, *55*(6), 1954-1956(1984)

21. Klaui, M. *et al.*, Domain wall motion induced by spin polarized currents in ferromagnetic ring structures. *Applied physics letters*, *83*(1), 105-107 (2003)

22. Parkin, S. S. P., Hayashi, M. & Thomas, L. Magnetic domain-wall racetrack memory. *Science* **320**, 190–194 (2008).

23. Fert, A. & Levy, P. A. Role of anisotropic exchange interactions in determining the properties of spin glasses. *Phys. Rev. Lett.* **44**, 1538–1541 (1980).

24. Gilbert, T. L. A phenomenological theory of damping in ferromagnetic materials. *Magnetics, IEEE Transactions on*, *40*(6), 3443-3449 (2004)

25. Supplementary information

26. Ferré J., *et al.,* Universal magnetic domain wall dynamics in the presence of weak disorder *C. R. Physique* **14** 651–666 (2013)

27. Chauve P., Giamarchi T., & Le Doussal P. Creep and depinning in disordered media *Phys. Rev. B* **62**, 6241 (2000)

28. Zakeri, K., Zhang, Y., Chuang, T. H., & Kirschner, J.. "Magnon lifetimes on the Fe (110) surface: The role of spin-orbit coupling" *Physical review letters,* 108(19), 197205. (2012)

29. Freimuth F., Blugel S., & Mokrousov Y. Berry phase theory of Dzyaloshinskii-Moriya interaction and spin-orbit torques *arXiv:1308.5983*

30. Kim, K. W., Lee, H. W., Lee, K. J., & Stiles, M. D. Chirality from interfacial spin-orbit coupling effects in magnetic bilayers. *Phys. Rev. Lett. 111*(21), 216601 (2013)

31. Gorchon, J., et al. "Pinning-Dependent Field-Driven Domain Wall Dynamics and Thermal Scaling in an Ultrathin Pt/Co/Pt Magnetic Film." *Phys. Rev. Lett.* 113.2  027205 (2014)

32. Allwood, D. A. *et al*. Magnetic domain-wall logic. *Science* **309**, 1688–1692 (2005).

**Acknowledgements**
We thank H. W. Lee, K-J Lee and T. Giamarchi for helpful discussions as well as A Thiaville, S Pizzini, and J Vogel for critically reading the manuscript and discussing the results. This work was partially supported by the ANR (11 BS10 008, ESPERADO) project and European Commission under the Seventh Framework Programme (GA 318144, SPOT) and (GA-2012-322369, Sport for Memory). AM has been supported by King Abdullah University of Science and Technology.

**Competing financial interests**
The authors declare no competing financial interests

# SUPPLEMENTARY INFORMATION

S1. Thickness dependence of DW motion asymmetry and interface anisotropy

S2. Prefactor of the DW creep scaling law

S3. Influence of damping on the DW velocity in the thermally activated regime

   a) **Chiral damping and thermal fluctuations**
   b) **Depinning time vs. Gilbert damping**
   c) **Velocity dependence on damping in the thermally activated regime**

S4. Asymmetry dependence on DW velocity.

S5. Link between the asymmetry and $m_{ip}$ variation

## S1. Thickness dependence of DW motion asymmetry and interface anisotropy.

To probe the dependence of this phenomenon on the material properties, we have compared the asymmetries measured on several samples having different thickness of the Pt upper layer (15Å–70Å). We chose this approach, to ensure that neither the magnetic properties of the Co layer nor the interface asymmetry[1,2,3], known to be essential for the DMI, are affected. All the samples exhibit the same qualitative behavior: $A$ varies up to $H_{sat}$, and fully saturates above. Surprisingly, $A$ decreases as the top Pt is made thicker. The possibility to experimentally turn off $A$ affords another independent way of probing its origin. The DMI creates an asymmetry by inducing a bias field that shifts the velocity curves towards positive $H_{ip}$ for one DW and towards negative $H_{ip}$ for the other (Figure 1c). Consequently, the saturation value of $A$ should be proportional to the bias field ($H_{DMI}$). And since $H_{sat}$ is the experimental indicator of $H_{DMI}$, the amplitude of $A$ should be related to $H_{sat}$. Experimentally we do not observe such a connection: the variations of $A$ are not associated to changes of $H_{sat}$ (Figure S1), thus showing that $A$ does not originate from $H_{DMI}$.

The uniaxial anisotropy fields ($H_k$) were determined from measurements of anomalous Hall effect for all the samples. The results plotted in figure S1 shows a reduction of the anisotropy fields from the sample with the thinnest (15.6Å) Pt capping to the thickest (70Å).

The variation of the anisotropy field reflects a change of the interfacial anisotropy. While $H_k$ varies by 30% (0.55 T to 0.38 T) after correcting for the layer's demagnetizing field ($\mu_0 M_s$=1.3 T), we calculate a maximum variation of the anisotropy constant of 10%.

Though the relative change of the interfacial anisotropy is small, it is nevertheless surprising to observe such a variation for Pt capping layers exceeding several nm. A plausible hypothesis is that the heating of the sample during deposition may influence its crystallographic structure in the vicinity to the interface. It is well known that high temperature annealing enhances the Pt and Co intermixing.

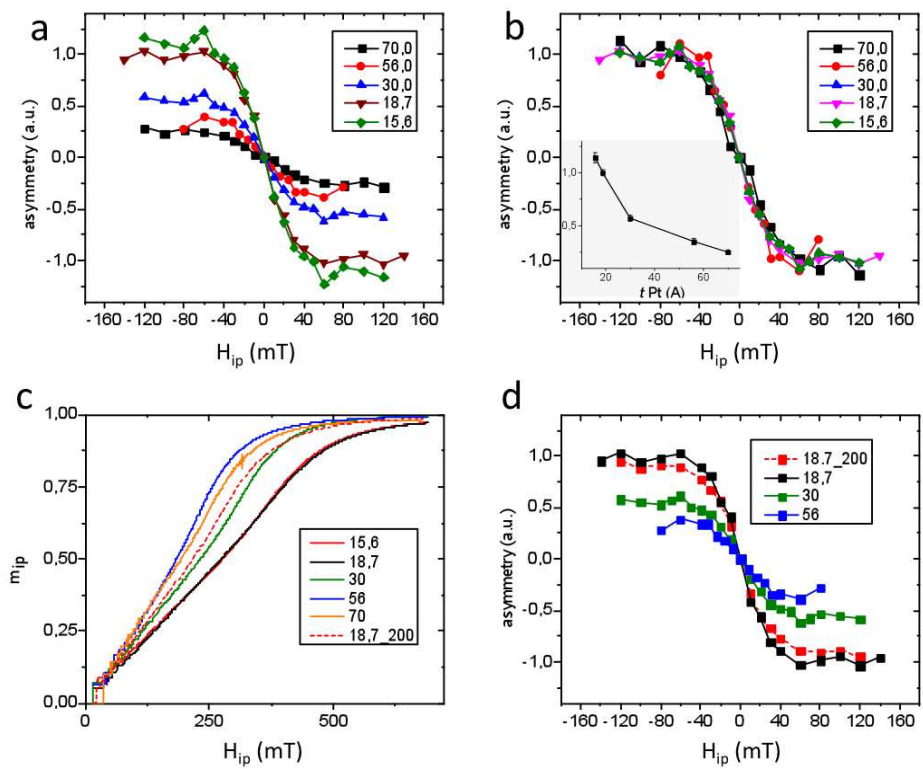

**Figure S1**. **a.** Asymmetry curves extracted from Kerr microscopy images for Pt thickness ranging from 15 Å to 70 Å. **b.** When normalized to 1 all the curves overlap, indicating that $H_{sat}$ is the same for all the samples. The inset shows the dependence of the normalizing factor (saturation value of the asymmetry) with the thickness of the Pt capping. **c.** Measurement of the anisotropy fields of all the samples used in the experiment. **d.** Asymmetry of the annealed sample compared to the as deposited ones. If the asymmetry value would be correlated to the anisotropy field the asymmetry of the annealed sample (red) would be comprised in-between the asymmetries measured for the samples with 30Å and 56Å of Pt capping.

In order to corroborate this hypothesis we have grown several series of the following samples: $Pt_{30}Co_6Pt_{20}$, $Pt_{30}Co_6Pt_{60}$ and $Pt_{30}Co_6Pt_{3520}$. While the $Pt_{60}$ layer of the second was deposited continuously, the Pt capping of the third was deposited in three steps consisting of $Pt_{20}$ separated by 30 seconds pauses, to allow cooling. **N.B.** The changes of $H_k$ correspond to small variations of the interfacial anisotropy constant, and therefore are sensitive to small variations of the growth conditions.

Despite the variations from series to series, we consistently find that $H_k(Pt_{30}Co_6Pt_{20}) > H_k(Pt_{30}Co_6Pt_{3520})+50mT > H_k(Pt_{30}Co_6Pt_{60})+100mT$, in agreement with the different heating that the samples might have suffered.

To verify experimentally that the heat-induced changes in the anisotropy do not affect the asymmetry we annealed the $t_{Pt}$=18.7Å sample at 200°C in vacuum for 30 minutes, and measured again its $H_k$ as well as the DW motion asymmetry. Indeed, we observe that the annealed sample has a smaller anisotropy field (the same as the sample with $t_{Pt}$=30Å), consistent with the Pt-Co inter-diffusing, while the asymmetry remains almost unchanged (Figure S1 d). The only consequence of the anisotropy variation is that the DW width might vary from sample to sample. However, since it depends on the square root of the anisotropy field, its variations will not exceed 10%. Consequently, the DW demagnetizing field stabilizing the Bloch DW structure will vary by the same ratio. In the absence of DMI, $H_{sat}$ is given by the DW demagnetizing field. The fact that we do not detect any significant sample-to-sample variation of $H_{sat}$ is consistent with this scenario.

## S2. The prefactor of the DW creep scaling law

Recently, Gorchon et al.[7] evidenced experimentally an exponential dependence of the creep velocity prefactor on temperature. They found that:

$$v_0 = d_0 f_0 \cdot \exp\left[ C \cdot \frac{U_c}{k_B T} \right]$$

$d_0 f_0$, is the product of the pinning potential correlation length ($d_0$) and the depinning attempt frequency ($f_0$). $C \approx 1$ is an empirical constant. Inserting this formula into the creep scaling law leads to:

$$v_{DW} = d_0 f_0 \cdot \exp\left[ -\frac{U_c H_p^{1/4}}{k_B T} \cdot \left( H_z^{-1/4} - H_c^{-1/4} \right) \right], \text{ where } H_c = \frac{H_p}{C^4}$$

By rewriting the creep law under this form all the energy contributions are removed from the velocity prefactor; they are included in $H_c^{-1/4}$, which is an offset of the abscissa.

Figure S2 represents schematically the changes of the creep scaling law produced by variations of the energy barrier and of the attempt frequency. Changes of the pinning barrier have two effects. On one hand, the slope is modified, but the curve is also shifted laterally, due to the variation of the pinning field (related to the energy barrier). On the contrary, the attempt frequency does not produce any variation of the slope; it only shifts the curve vertically. Our experimental data (Figure 3 in the main text) does not exhibit any significant change of slope when $H_{ip}$ is reversed (from -80mT to 80mT); we only observe a shift. We deduce that this must be a vertical shift (associated to $f_0$), since a lateral shift (associated to the pinning field) would be accompanied by a change of slope. The counter example can be seen when comparing the curve at $H_{ip}$=0mT to the curves obtained at 80mT. Their difference in slope is associated to a change of the pinning field (Figure 3 in the main text).

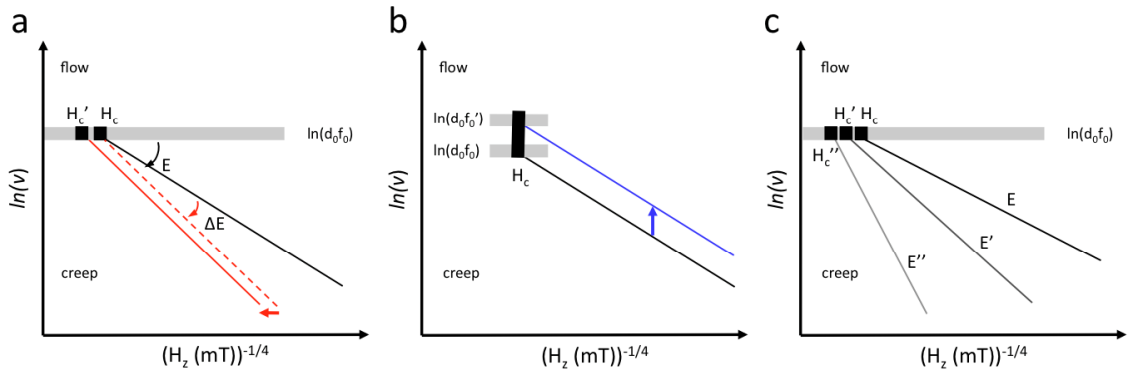

**Figure S2**. Influence of energy barrier and attempt frequency on the DW creep scaling law. **a.** The variations of the pinning barrier lead to a change of slope (red dotted line), but also to a lateral shift due to a change of the pinning field. The horizontal gray line represents the approximate transition from creep to flow. The black square is the approximate $H_c^{-1/4}$ **b.** The variation of the attempt frequency leads only to a vertical shift of the curve. **c.** $H_c^{-1/4}$ and $d_0 f_0$ can be roughly estimated from the intercept of the creep velocity curves corresponding to different pinning barriers.

*Order of magnitude of the attempt frequency.*

Besides the measurements of Gorchon et al.[7] the order of magnitude of $d_0 f_0$ can be extracted also from previous measurements of DW creep in Pt/Co/Pt samples with varying Co thickness[6]. In this case it is the change of thickness that leads to changes of the effective parameters (exchange, anisotropy, pinning…) affecting the energy landscape. Figure 2d of supplementary reference 6, exhibiting the DW velocity in the $\ln(v_{DW}) vs. H_z^{-1/4}$ format, shows that all the curves corresponding to the different samples converge close to a small area. This area allows estimating the range of $H_c^{-1/4}$ and $d_0 f_0$ values for the series of samples. Since our samples are nominally identical, and are made under the same conditions in the same sputtering chamber as these previous studies[6,7], we use their estimate for the order of magnitude of $d_0 f_0$. Independent estimates from these two studies coincide to $d_0 f_0 \approx 10 m/s$. Considering this value, we estimate that the deviations from the creep regime should occur when $\ln(v_{DW}) \approx 2.5$. By considering that the disorder correlation length should be larger than the typical grain size (20 nm), the order of magnitude of the attempt frequency could be anywhere between 10MHz and 1GHz.

## S3. Influence of damping on the DW velocity in the thermally activated regime

Thermally activated DW motion consists of relatively fast DW displacements interrupted by pinning and de-pinning events. If the driving field is sufficiently small compared to the pinning field, the DW spends most of the time trapped at pinning sites, trying to de-pin. When the temperature fluctuations finally free it, the DW moves to the next pinning center. In this motion regime the experimentally determined DW velocity is much smaller than the steady DW velocity (which is effective in-between pinning events); it is merely a measure of the average de-pinning time. For this reason, instead of trying to assess the complete influence of damping on the thermally activated DW velocity we will first focus on the DW depinning.

The thermally activated de-pinning of a DW depends on two ingredients: the usually dominant one is the exponential dependence on the height of the energy barrier in units of thermal energy. The other, often neglected, is the pre-factor of this exponential, given by the attempt frequency ($p = f_0 \cdot \exp(\Delta E / k_B T)$). While the energy barrier does not depend on the damping value, the attempt frequency may be altered by variations of the damping. We analyze this possibility numerically.

### a) Chiral damping and thermal fluctuations

The effect of temperature along with the chiral damping proves to be impossible to include in the numerical modeling. We find that the temperature alone induces DW displacements. This is a direct consequence of a combination of two factors:
1) the random thermal field is created using a pseudo random number generator;
2) the chiral damping makes the DW behave as a ratchet.

A ratchet is an object having an asymmetric response to excitations[4,5]. When subjected to periodic excitations a ratchet produces a net directional motion (rotation, displacement, etc). In the case of DW motion, the asymmetric response is given by the chiral damping. And since the temperature fluctuations are generated using a pseudo random number generator whose autocorrelation is not null, it includes periodic components that inevitably create an artificial displacement.

Despite this difficulty, it is still possible to test the effect of damping in the thermally activated DW motion, using an alternative approach. As long as $m_{ip}$ is established by the competition between $H_{ip}$ and the DW's demagnetizing field, the only variable determining the damping value is $H_{ip}$. Therefore, instead of computing the velocity dependence on $H_{ip}$ we can use the damping value as the free parameter. We use a constant damping (independent of $m_{ip}$) and vary its value. These two methods are equivalent as long as $H_z$ is sufficiently small, not to induce any supplementary DW deformation and $m_{ip}$ is determined only by the balance between $H_{ip}$ and the DW's demagnetizing field. This approximation is valid in the case of very slow thermally activated DW motion (corresponding to our experimental conditions).

**b) Depinning time vs. Gilbert damping**

In the micromagnetic simulations (Figure 3d of the main text) we observed a net dependence of the depinning time on the damping coefficient. Since the energy barrier for the two simulations is the same, this difference may only originate in the DW's attempt rate. Furthermore, the fact that the depinning time is proportional to the damping value shows that the DW mobility plays a significant role in the motion of the DW inside the pinning center. The DW depinning is a probabilistic process whose full characterization requires averaging over a statistically significant amount of events. Consequently, micromagnetic modeling, being relatively time consuming, is not the best-suited tool for the extensive study of the thermally activated DW motion. In order to further analyze the influence of the damping coefficient on the DW depinning we use a numerical collective coordinate model.

**c) Velocity dependence on damping in the thermally activated regime**

While 2D micromagnetics allow including realistically the partial pinning of the DW, by definition in the 1D models the entire DW is enclosed in the pinning potential. As a consequence, when projecting the 2D DW dynamics onto a single dimension, the effective pinning potential needs to be adapted accordingly. The simplest way is to use effective pinning centers that are much wider than the DW, such that the DW will have to shift far from the energy minimum before actually applying pressure on the pinning potential.

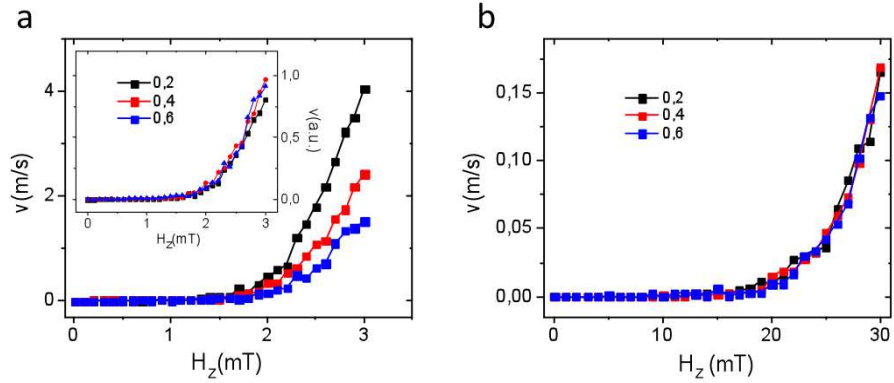

**Figure S3**. DW velocity in the 1D model for 3 different damping values. Each velocity point is obtained after averaging 30 displacements produced by 500ns long pulses **a.** DW velocity when the width of the pinning potential is 10 times larger than the size of the DW (5nm). The inset shows the velocity curves normalized by the inverse of the damping value. **b.** The velocity calculated for a potential half the width of the DW (5nm). In this case, the DW velocity is independent of the damping value.

In order to probe the influence of the damping variation on the thermally activated DW motion, we calculate the velocity (Figure S3) as a function of the perpendicular magnetic field, for several damping values. The DW parameters used in the simulations are $\Delta DW=5nm$, $H_{dip}=40mT$. The effective pinning field is modeled by a sinusoidal function with a period of 50nm, (10 times larger than the DW width) with amplitude of 3 mT. This form of pinning is chosen only for ease of implementing in the numerical model. Though it is not realistic for the quantitative calculation of the DW velocity, since it allows modeling the repeated pinning and de-pinning of the DW, it is sufficient to evidence the impact of damping on the DW velocity. We observe that the DW velocity depends strongly on the damping value (0.2, 0.4, 0.6). Moreover, by normalizing all the velocity curves by the inverse of the damping (proportional to the DW mobility) they overlap (figure S2a). This indicates that the DW velocity depends on the intrinsic DW mobility. On the contrary, when trapping the DW in a narrow potential well with the same barrier height (period of 2.5 nm and amplitude of pinning field of 60 mT) the velocity curves completely loose the damping dependence (figure S2b).

### d) Attempt frequency in the 1D model

To identify the physical parameter responsible for the damping dependence of the DW velocity, we analyze the motion of the DW driven only by temperature (without $H_z$) inside the pinning potential. To estimate the DW attempt frequency we compute the frequency spectrum (by FFT analysis) of the DW position. We observe that the attempt frequency is not described by a well-defined peak, but rather by a cut-off value.

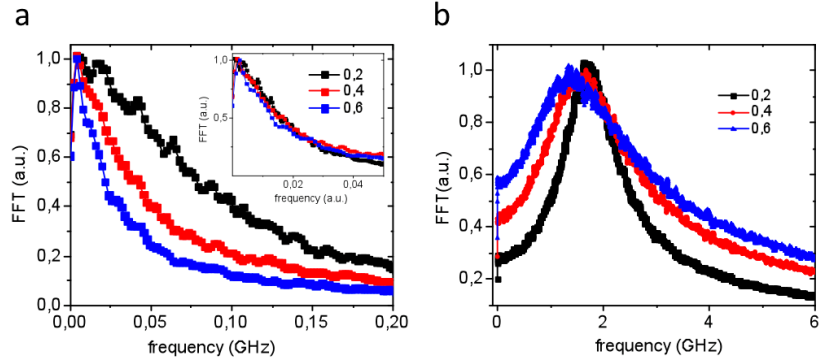

**Figure S4**. Fourier analysis of a pinned DW with Hz=0. The FFTs were obtained by averaging 100 independent evolutions with a duration of 500 ns each **a.** The case of a wide pinning potential. The inset shows that by normalizing the frequency by the domain wall mobility (inverse of the damping value), the curves overlap. **b.** For a narrow pinning potential the FFT exhibits a peak whose position is almost independent of damping.

When normalizing the different FFTs to the inverse of the damping (DW mobility) the spectra overlap. This proves that the characteristic attempt frequency of the DW (cut-off frequency) changes proportionally to its mobility. In the case where the pinning potential is not sufficiently wide (of the order of the DW width or smaller) the FFT shows a relatively well-defined resonance peak (corresponding to the DW's ferromagnetic resonance) whose position does not depend on damping. The fact that the attempt frequency (in this case given by the position of the peak) loses its damping dependence, explains why the DW velocity in the case of narrow pinning potential does not depend on damping.

## S4. Asymmetry dependence on DW velocity.

The numerical analysis of the DW motion reveals an important feature: since the velocity variation induced by the damping does not depend of the value of $H_z$ (Figure S2a), the measured asymmetry should also be independent of $H_z$.

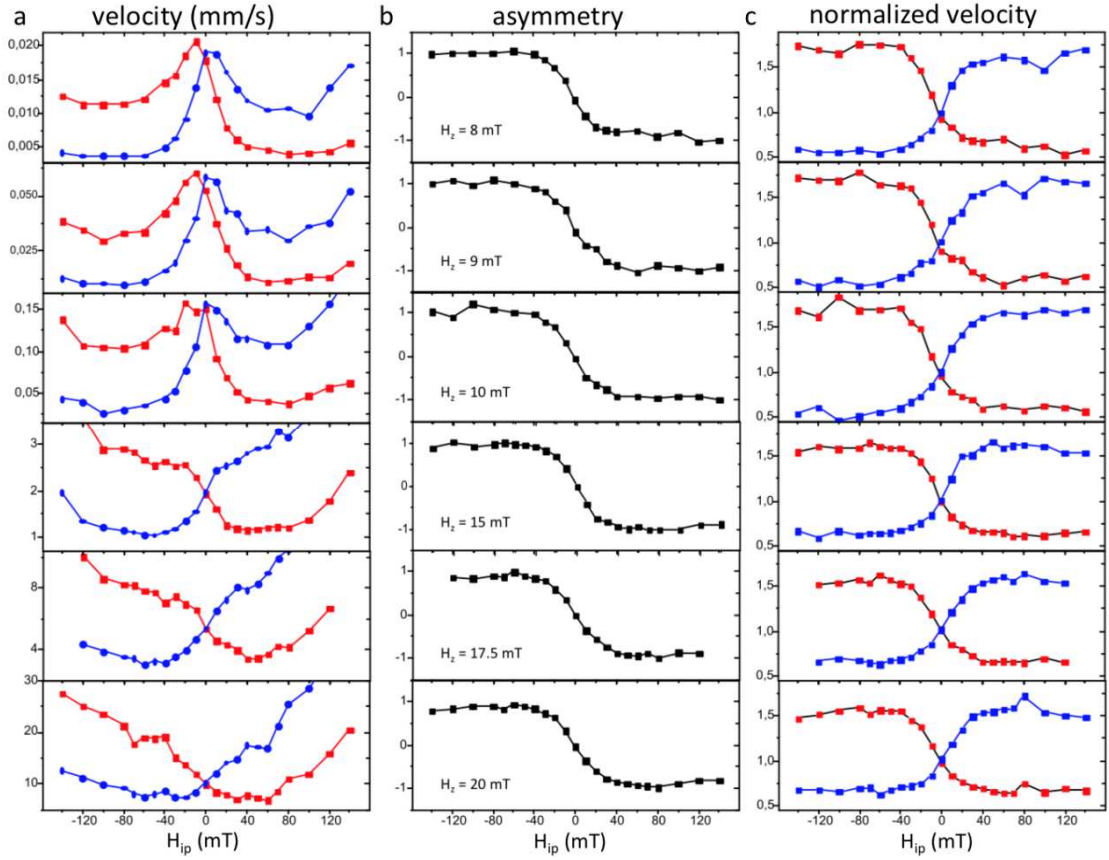

**Figure S5**. Asymmetry dependence on $H_z$. **a.** Velocity curves for different values of $H_z$. **b.** The corresponding asymmetries $A_{raw} = 2(v\downarrow\uparrow - v\uparrow\downarrow)/(v\downarrow\uparrow + v\uparrow\downarrow)$. **c.** The normalized velocities.

To verify this prediction, we repeated the measurements for different values of $H_z$. Indeed we find that, though the velocity changes by three orders of magnitude, both $H_{sat}$ and *A* remain the same. On the contrary, the symmetric component varies significantly, changing the overall shape of the curves. The uncorrelated behavior of the symmetric and asymmetric components of the velocity points to their different origin: while the asymmetry is given by the chiral damping through the attempt frequency, the symmetric part is linked to changes in the pinning potential induced by $H_{ip}$.

## S5. Link between the asymmetry and $m_{ip}$ variation

The DW velocity measurements correspond to the creep regime, where the DW velocity depends linearly on the attempt frequency and exponentially on the DW energy and pinning strength. For simplicity, we merge all energy contributions under a same parameter (E)

$v \approx f_0 \cdot \exp(E)$

We can write explicitly for the two DWs: $v_{\uparrow\downarrow} \approx f_{0\uparrow\downarrow} \cdot \exp(E_{\uparrow\downarrow})$ and $v_{\downarrow\uparrow} \approx f_{0\downarrow\uparrow} \cdot \exp(E_{\downarrow\uparrow})$

We observe that below $H_{sat}$ the two DW velocities vary differently, while above, their ratio stays constant.

Since $\dfrac{v_{\uparrow\downarrow}}{v_{\downarrow\uparrow}} = \dfrac{f_{0\uparrow\downarrow}}{f_{0\downarrow\uparrow}} \cdot \exp(E_{\uparrow\downarrow} - E_{\downarrow\uparrow})$

$v_{\uparrow\downarrow}/v_{\downarrow\uparrow} = cst$ implies that both $f_{0\uparrow\downarrow}/f_{0\downarrow\uparrow} = cst$ and $E_{\uparrow\downarrow} - E_{\downarrow\uparrow} = cst$, or that the attempt frequency and energy variations coincidently compensate each other. This scenario can be excluded based on the robustness of the observation: the asymmetry saturation is observed for a wide range of $H_{ip}$ for all our samples. There are two possibilities left:

Case 1) Damping induced asymmetry

This mechanism links directly the two DW velocities to $m_{ip}$. Therefore, as indicated in the manuscript the asymmetry saturation coincides with the saturation of the DWs magnetization.

Case 2) energy induced asymmetry

In thin films with vertical SIA and isotropic in the plane, the chiral energy for the two DWs is:

$$E \propto \int_{\Delta_{DW}} \frac{\partial \theta}{\partial x} \cos(\varphi) dx$$

The three parameters that define this energy and that can be in principle affected by $H_{ip}$ are: the magnetization gradient, the DW width and the $m_{ip}$ orientation. Since the variation and the saturation of the asymmetry occur at low fields, while the magnetization gradient varies significantly at much larger fields, comparable with the uniaxial anisotropy, we deduce that the chiral energy variation is essentially due to a variation of $m_{ip}$.

$$E_{\uparrow\downarrow} - E_{\downarrow\uparrow} \propto \int_{\Delta_{DW}} \frac{\partial \theta}{\partial x} \cos\varphi_1(H_{ip}) dx - \int_{\Delta_{DW}} \frac{\partial \theta}{\partial x} \cos\varphi_2(H_{ip}) dx$$

Since $\dfrac{\partial \theta}{\partial x}$ has opposite sign for up/down and down/up DWs, and $\cos(\varphi)$ variations induced by $H_{ip}$ have the same sign, any variation of $m_{ip}$ for any one of the DWs will amplify $E_{\uparrow\downarrow} - E_{\downarrow\uparrow}$. As a consequence, $E_{\uparrow\downarrow} - E_{\downarrow\uparrow} = cst$ is verified only if $m_{ip}$ is constant for both DWs.

Therefore, independently of the mechanism that creates the asymmetry, its saturation at low fields, is a robust indicator of the $m_{ip}$ saturation.

Supplementary References:

31. Bode, M. *et al*. Chiral magnetic order at surfaces driven by inversion asymmetry. *Nature* **447**, 190–193 (2007).
32. Je, S-G et al. Asymmetric Magnetic Domain-Wall Motion by the Dzyaloshinskii-Moriya Interaction Physical Review B 88, 214401 (2013)
33. Thiaville, A., Rohart, S., Jué, É., Cros, V., & Fert, A. Dynamics of Dzyaloshinskii domain walls in ultrathin magnetic films. *EPL (Europhysics Letters)*, *100*(5), 57002. (2012).
34. Franken, J. H., Swagten, H. J. M., & Koopmans, B. (2012). Shift registers based on magnetic domain wall ratchets with perpendicular anisotropy. *Nature Nanotechnology*, *7*(8), 499-503
35. Pérez-Junquera, A. *et al.,* Crossed-ratchet effects for magnetic domain wall motion. *Physical review letters*, *100*(3), 037203 (2008)
36. Metaxas, P. J., et al. "Creep and flow regimes of magnetic domain-wall motion in ultrathin Pt/Co/Pt films with perpendicular anisotropy." Phys. Rev. Lett. 99.21 (2007): 217208.
37. Gorchon, J., et al. "Pinning-Dependent Field-Driven Domain Wall Dynamics and Thermal Scaling in an Ultrathin Pt/Co/Pt Magnetic Film." *Phys. Rev. Lett.* 113.2 027205 (2014)